\documentclass[11pt]{article}
\usepackage{hyperref}
\usepackage{bm}
\usepackage{datetime}
\begin{document}

\begin{center}
\large{\bf Supersonic Jet Excitation using Flapping Injection}
\\\vspace{1cm}
Haukur Hafsteinsson$^{1}$, Lars-Erik Eriksson$^{1}$, Niklas Andersson$^{1}$ \\
Daniel Cuppoletti$^{2}$, Ephraim Gutmark$^{2}$ \\
Erik Prisell$^{3}$\\
\vspace{6pt}
\footnotesize{$^{1}$Chalmers University of Technology, Department of Applied Mechanics,\\ Gothenburg, Sweden}\\
\footnotesize{$^{2}$University of Cincinnati, Gas Dynamics and Propulsion Laboratory,\\ Cincinnati, Ohio}\\
\footnotesize{$^{3}$Swedish Defence Material Administration, Aero Propulsion and Power,\\ Stockholm, Sweden}
\end{center}

\begin{abstract}
Supersonic jet noise reduction is important for high speed military aircraft. Lower acoustic levels would reduce structural fatigue leading to longer lifetime of the jet aircraft.
It is not solely structural aspects which are of importance, health issues of the pilot and the airfield personnel are also very important, as high acoustic levels may result in severe hearing damage. It remains a major challenge to reduce the overall noise levels of the aircraft, where the supersonic exhaust is the main noise source for near ground operation. Fluidic injection into the supersonic jet at the nozzle exhaust has been shown as a promising method for noise reduction. It has been shown to speed up the mixing process of the main jet, hence reducing the kinetic energy level of the jet and the power of the total acoustic radiation. Furthermore, the interaction mechanism between the fluidic injection and the shock structure in the jet exhaust plays a crucial role in the total noise radiation. In this study, LES is used to investigate the change in flow structures of a supersonic (M=1.56) jet from a converging-diverging nozzle. Six fluidic actuators, evenly distributed around the nozzle exit, inject air in a radial direction towards the main flow axis with a total mass flow ratio of 3\%. Steady injection is compared with flapping injection. With flapping injection turned on, the injection angle of each injector is varied sinusoidally in the nozzle exit plane and the variation is the same for all injectors. This fluid dynamics video is submitted to the APS DFD Gallery of Fluid Motion 2013 at the 66 the Annual Meeting of the American Physical Society, Division of Fluid Dynamics (24-26 November, Pittsburgh, PA, USA).

\end{abstract}
\section*{Video Description}
First, a general picture is brought up to make the audience acquainted with the application. A simplified sharp throat converging-diverging nozzle in a model scale, is attached to a full size aircraft to show its actual location in real a application. Then, a slice through the full three-dimensional computational domain is showed. The domain reaches approximately 70 nozzle exit diameters downstream of the nozzle exit plane and about 4 nozzle exit diameters in the upstream direction. The flow field is obtained by solving the compressible Navier-Stokes equations using an in-house finite volume LES solver based on the G3D family of codes originally developed by Eriksson \cite{eriksson:95}. The computational grid used for the simulations consists of approximately 20 million cells and the simulations are done on 80 CPU's using MPI. Three cases are shown in the video; first, the baseline supersonic case without injection is shown, second a case with a steady injection at the nozzle exit is shown and finally a case with a flapping injection. For all three cases, the nozzle is operated at a nozzle pressure ratio (NPR) of 4.0, which gives a jet-exit Mach number of $M=1.56$. For all the cases a slice through the domain colored by $\nabla^2 p$ is showed. This quantity effectively shows sudden spatial pressure variations, such as those that occur across shocks. Inside the nozzle a stationary supersonic flow field is formed. It can be noticed that a shock is formed at the sharp throat. This is a conical shock which reaches further downstream, reflects on its self towards the nozzle wall at the jet center axis. Upon reaching the nozzle wall it reflects again this time passing through the nozzle exit towards the jet center axis where it reflects radially outwards. When interacting with the shear layer it reflects back again towards the jet center axis and so on generating a set of quasi-stationary compression- and expansion waves within the jet plume. Another similar set of shocks is formed at the nozzle lip. These two shocks generate a double shock cell structure which dissipates downstream. Plotting the $M=1.0$ iso-contour reveilles the location of the boundary between the supersonic jet-core region and its subsonic surroundings. A close look at the nozzle exit shows a relatively stable shear layer which quickly unfolds into circumferential vortex cores, as the flow transitions to high turbulence levels due to steep axial velocity gradient in the radial direction. 

The steady-state injection consists of 6 evenly distributed actuators around the nozzle exit. The injection angle is normal to the nozzle inner wall and is therefore directed radially inward towards the jet center axis. The total mass flow of all six injectors ($\dot{m}_\mathrm{i}$) compared to the mass flow through the nozzle throat  ($\dot{m}_\mathrm{j}$) is $\dot{m}_\mathrm{i}/\dot{m}_\mathrm{j}=3\,\%$, which is considered as relatively high mass flow for practical applications. The injection has a profound effect on the jet dynamics. The $M=1$ iso-surface shows how the fluidic injectors penetrate into the shear layer and create axial vortices which are convected downstream by the main jet flow. These vortices result in increased mixing of the jet plume with its ambient air and hence the length of the potential core is reduced. Since the injectors penetrate rather deepl into the jet, the main jet flow senses a blockage and its path is forced in between the injectors. Therefore, the radial location of the $M=1$ iso-contour increases in between the injectors. Thereafter, iso-contours of $\nabla^2 p$ is showed to visualize spectacular stationary bow-shocks formations upstream of each injector.

Continuing from the animation of the stationary bow-shocks formed due to the steady injection, the injection is switched to a flapping mode which shows how the position of the bow-shocks start to follow the flapping angle. The flapping amplitude is $\pm 60^{\circ}$, the frequency of the flapping is $f=1000\mathrm{Hz}$ and as mentioned earlier, the flapping angle is in the nozzle exit plane. Shifting back to a view of the $M=1$ iso-contour, shows how the mixing-rate is dramatically increased as the flapping injectors introduce a spinning motion to the shear layer. Furthermore, looking at the slice-through the domain showing $\nabla^2 p$ along the jet axis, an interesting motion of the double shock cell structure may be noticed. The flapping injection introduces a shock motion which can be referred to as ``shock pumping movement'', i.e. the two shock-cells keep more or less their original structure but the distance between them shifts back and forth. This results in a constructive and destructive shock superposition. This phenomenon is thought to be responsible for strong undesirable acoustic screech harmonics observed in the far-field as shown by Hafsteinsson et al. \cite{Hafsteinsson}.

A final view along the jet-axis towards the nozzle exit, shows how the flapping injection creates stunning spinning-shock formations with highly complex three dimensional shock interactions.

%

\end{document}